\documentclass[12pt,preprint]{aastex}

\usepackage{enumerate}
\usepackage[lofdepth,lotdepth,caption=false]{subfig}
\usepackage{gensymb}
\usepackage{textcomp}
\usepackage[utf8]{inputenc}
\usepackage[T1]{fontenc}
\usepackage{graphicx}

\shorttitle{MWA observations of the Moon}
\shortauthors{McKinley, Briggs, Kaplan et al}

\begin{document}

\title{Low Frequency Observations of the Moon \\ with the Murchison Widefield Array}

\def\ASU{$^{5}$}
\def\ANU{$^{1}$}
\def\CSIRO{$^{11}$}
\def\Curtin{$^{6}$}
\def\CfA{$^{4}$}
\def\NCRA{$^{8}$}
\def\Haystack{$^{12}$}
\def\MIT{$^{14}$}
\def\RRI{$^{13}$}
\def\USwinburne{$^{10}$}
\def\UMelbourne{$^{16}$}
\def\USydney{$^{7}$}
\def\UTasmania{$^{17}$}
\def\UW{$^{15}$}
\def\UWA{$^{18}$}
\def\UWisc{$^{3}$}
\def\Victoria{$^{9}$}
\def\CAASTRO{$^{2}$}

\author{
B.~McKinley\ANU$^,$\CAASTRO, 
F.~Briggs\ANU$^,$\CAASTRO, 
D.~L.~Kaplan\UWisc, 
L.~J.~Greenhill\CfA,
G.~Bernardi\CfA, 
J.~D.~Bowman\ASU, 
A.~de Oliveira-Costa\CfA,
S.~J.~Tingay\Curtin$^,$\CAASTRO, 
B.~M.~Gaensler\USydney$^,$\CAASTRO, 
D.~Oberoi\NCRA, 
M.~Johnston-Hollitt\Victoria,
W.~Arcus\Curtin, 
D.~Barnes\USwinburne, 
J.~D.~Bunton\CSIRO, 
R.~C.~Cappallo\Haystack, 
B.~E.~Corey\Haystack, 
A.~Deshpande\RRI, 
L.~deSouza\CSIRO$^,$\USydney,
D.~Emrich\Curtin,
R.~Goeke\MIT,
B.~J.~Hazelton\UW, 
D.~Herne\Curtin, 
J.~N.~Hewitt\MIT, 
J.~C.~Kasper\CfA, 
B.~B.~Kincaid\Haystack, 
R.~Koenig\CSIRO, 
E.~Kratzenberg\Haystack, 
C.~J.~Lonsdale\Haystack, 
M.~J.~Lynch\Curtin, 
S.~R.~McWhirter\Haystack,
D.~A.~Mitchell\UMelbourne$^,$\CAASTRO, 
M.~F.~Morales\UW, 
E.~Morgan\MIT, 
S.~M.~Ord\Curtin$^,$\CfA, 
J.~Pathikulangara\CSIRO, 
T.~Prabu\RRI, 
R.~A.~Remillard\MIT, 
A.~E.~E.~Rogers\Haystack, 
A.~Roshi\RRI, 
J.~E.~Salah\Haystack, 
R.~J.~Sault\UMelbourne, 
N.~Udaya~Shankar\RRI, 
K.~S.~Srivani\RRI, 
J.~Stevens\CSIRO$^,$\UTasmania, 
R.~Subrahmanyan\RRI$^,$\CAASTRO, 
R.~B.~Wayth\Curtin$^,$\CAASTRO, 
M.~Waterson\Curtin$^,$\ANU,
R.~L.~Webster\UMelbourne$^,$\CAASTRO, 
A.~R.~Whitney\Haystack, 
A.~Williams\UWA, 
C.~L.~Williams\MIT, 
J.~S.~B.~Wyithe\UMelbourne$^,$\CAASTRO
\\ 
$^{1}$The Australian National University, Canberra, Australia\\
$^{2}$ARC Centre of Excellence for All-sky Astrophysics (CAASTRO)\\
$^{3}$University of Wisconsin--Milwaukee, Milwaukee, USA\\
$^{4}$Harvard-Smithsonian Center for Astrophysics, Cambridge, USA\\
$^{5}$Arizona State University, Tempe, USA\\
$^{6}$International Centre for Radio Astronomy Research, Curtin University, Perth, Australia\\
$^{7}$University of Sydney, Sydney, Australia\\
$^{8}$National Centre for Radio Astrophysics, Tata Institute for Fundamental Research, Pune, India\\
$^{9}$Victoria University of Wellington, New Zealand\\
$^{10}$Swinburne University of Technology, Melbourne, Australia\\
$^{11}$CSIRO Astronomy and Space Science, Australia\\
$^{12}$MIT Haystack Observatory, Westford, USA\\
$^{13}$Raman Research Institute, Bangalore, India\\
$^{14}$MIT Kavli Institute for Astrophysics and Space Research, Cambridge, USA\\
$^{15}$University of Washington, Seattle, USA\\
$^{16}$The University of Melbourne, Melbourne, Australia\\
$^{17}$University of Tasmania, Hobart, Australia\\
$^{18}$University of Western Australia, Perth, Australia\\
}

\begin{abstract}
A new generation of low frequency radio telescopes is seeking to observe the redshifted 21 cm signal from the Epoch of Reionization (EoR), requiring innovative methods of calibration and imaging to overcome the difficulties of widefield low frequency radio interferometry. Precise calibration will be required to separate the small expected EoR signal from the strong foreground emission at the frequencies of interest between 80 and 300 MHz. The Moon may be useful as a calibration source for detection of the EoR signature, as it should have a smooth and predictable thermal spectrum across the frequency band of interest. Initial observations of the Moon with the Murchison Widefield Array 32 tile prototype show that the Moon does exhibit a similar trend to that expected for a cool thermally emitting body in the observed frequency range, but that the spectrum is corrupted by reflected radio emission from Earth. In particular, there is an abrupt increase in the observed flux density of the Moon within the internationally recognised Frequency Modulated (FM) radio band. The observations have implications for future low frequency surveys and EoR detection experiments that will need to take this reflected emission from the Moon into account. The results also allow us to estimate the equivalent isotropic power emitted by the Earth in the FM band and to determine how bright the Earth might appear at metre wavelengths to an observer beyond our own solar system.
\end{abstract}



\section{Introduction}
The astrophysics of the Intergalactic Medium over the redshift interval z $\approx$ 200 to 8 is expected to give rise to observable phenomena in the redshifted 21 cm line of neutral hydrogen (see e.g. \citet{pritchard_loeb,moraleswyithe2010,furlanetto2006} for detailed reviews). The 21 cm line from this period, which extends from the Dark Ages through the Epoch of Reionization (EoR), redshifts to radio frequencies of between 30 and 200 MHz, making it a suitable target for the next generation of low frequency radio telescopes. Use of the redshifted 21cm line of neutral hydrogen to study the EoR will require exquisite calibration in order to achieve milli-Kelvin level spectral line sensitivity in the presence of Galactic and extragalactic foregrounds that are brighter than the expected signal by at least 2 to 3 orders of magnitude \citep{shaver_etal_1999, rogersbowman2008, berardi2009}. One suggestion \citep{shaver_etal_1999} for instrumental calibration for observation of the global EoR signature \citep{bowmanrogers2009} would be to use the smooth, thermal spectrum of the Moon for comparison with the sky brightness.

The goals of the work reported here are to assess the suitability of the Moon as a comparison source for EoR detection and to test novel observing modes for the Murchison Widefield Array (MWA) \citep{tingay,mwadesign} that would span the full range of the uncertainty in redshift that surrounds the reionization phase change. 

A concern at the outset has been that the lunar spectrum could be corrupted by reflected communications and navigation transmissions from the Earth. It has been known since the US Army Signal Corps' Project Diana \citep{projdiana} that radio signals could be reflected by the Moon and received on Earth. Indeed, it is now common practice for amateur radio enthusiasts to bounce communications signals off the moon \citep{katz} to be received by listeners on the other side of the planet. Radio leakage from the Earth has been modelled theoretically \citep{sullivan1978} and measured by observing the reflected signals off the Moon using the Arecibo telescope \citep{sullivan_moon}. However, the use of radio communications on Earth has changed since these studies were undertaken and the EoR waveband of 80 to 200 MHz covered by the MWA deserves an updated study and continued monitoring. A strength of the present observations with the MWA 32 tile prototype (MWA 32T) is that the gain of the synthesised beam isolates the lunar reflections from any ambient signals that may enter weakly into the radio quiet zone of the Murchison Radio Observatory (MRO) on more direct paths from local transmitters.

At the frequencies of interest the radio sky is bright and dominated by synchrotron radiation from the Milky Way. The sky background varies as a power law function of frequency \citep{rogersbowman2008,deoliveiracosta2008} and the sky brightness temperature is given by:
\begin{equation} 
T_{sky}=T_{150}\left(\frac{\nu}{150 \ \rm{MHz}}\right)^{-\beta} \ \rm{K}
\label{skytemp}
\end{equation}
where the spectral index $\beta$ at high galactic latitudes has been measured \citep{rogersbowman2008} to be 2.5 and at the sky position of these observations (J2000 coordinates 0\textsuperscript{h}51\textsuperscript{m}6\textsuperscript{s}, 11\degree 24$'$40$''$) the sky temperature $T_{150}$ is 248 K \citep{landecker1970}.

The brightness temperature of the Moon can be modelled by a combination of three factors; thermal emission, assuming a Moon brightness temperature of 230 K \citep{heilesdrake63}, reflected radio sky background emission, assuming a Moon reflectivity of 7\% \citep{evans_moon} and sky temperature given by equation (\ref{skytemp}), and reflected emission from the Earth. Reflected emission from the Sun, even at Full Moon, is expected to contribute less than 1K for a quiet Sun \citep{hagfors69}. This reflected solar emission is not included since, based on the information available from the National Oceanic and Atmospheric Administration Space Weather Prediction Center (NOAA SWPC), the solar activity during our observing period was low. The contribution to the brightness temperature from reflected emission from the Earth is left as an unknown parameter in this model as it is not well understood at these frequencies and for recent times. These observations aim to contribute to the characterisation of this reflected emission.

\begin{figure}[hbtp] 
\centering 
\includegraphics[]{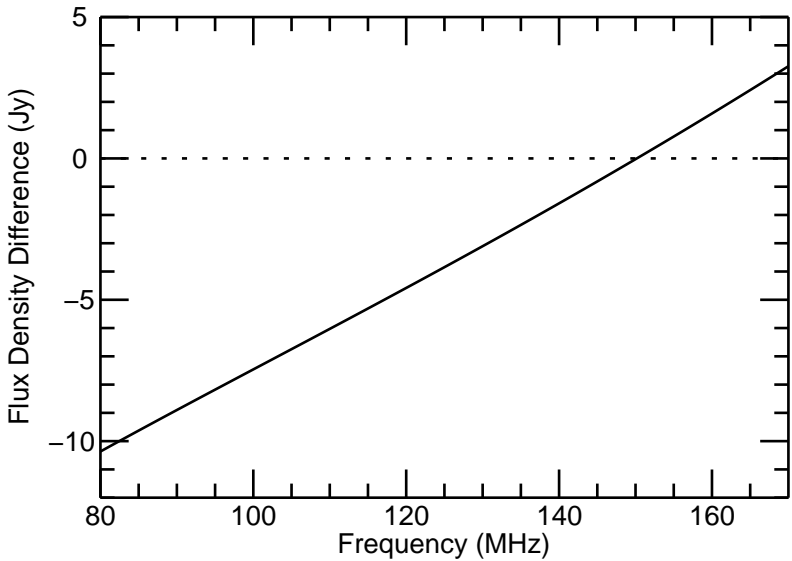}
\caption{Theoretical difference between Moon and sky flux density at J2000 coordinates 0\textsuperscript{h}51\textsuperscript{m}6\textsuperscript{s}, 11\degree 24$'$40$''$, for observations at 0030-0132 LST on UTC dates 24 and 25 September 2010 respectively.} 
\label{MoonSkyTheory}
\end{figure}

Brightness temperatures (T$_{b}$ in K) are converted to flux densities (S in Jy) by:
\begin{equation} 
S=\frac{2 k T_{b} \Omega}{\lambda^{2} 10^{-26}}
\label{tempconvert}
\end{equation}
where \emph{k} is the Boltzmann constant in J K$^{-1}$, $\Omega$ is the solid angle subtended on the sky in steradians and $\lambda$ is wavelength in m. The value of $\Omega$ used is $6.0 \times 10^{-5}$ which corresponds to the Moon's angular diameter of 30.0 arcminutes at the time of these observations. The expected difference in flux density between the Moon and the sky, ignoring reflected emission from Earth, is plotted in Figure \ref{MoonSkyTheory} and shows that at the lower end of the frequency band the Moon should appear as a \textquoteleft hole in the sky\textquoteright, crossing over to being a positive source relative to the sky background at around 150 MHz. Recovering this Moon-sky difference signal was an important aim of the MWA observations described in this paper.

\section{Observations}

The MWA is a new type of radio interferometer array that is designed to observe the redshifted 21 cm signal at frequencies of between 80 and 300 MHz, corresponding to a redshift range of z=4 to 17 \citep{tingay,mwadesign}. It consists of a large number of small, crossed-dipole antennas grouped into \textquoteleft tiles\textquoteright \ whose beams are electronically steered using analog beamformers. Each tile is made up of 16 dipole antennas arranged in a 4 by 4 grid above a conducting mesh ground plane. A 32 tile prototype instrument (MWA 32T) was operated over the period September 2007 to September 2011 within the radio-quiet MRO located in outback Western Australia. More detailed descriptions of the MWA 32T prototype configuration can be found in \citet{oberoi}, \citet{ord} and  \citet{williams}. 

\begin{figure}[hbtp] 
\centering 
\includegraphics[width=1.0\textwidth,totalheight=0.9\textheight]{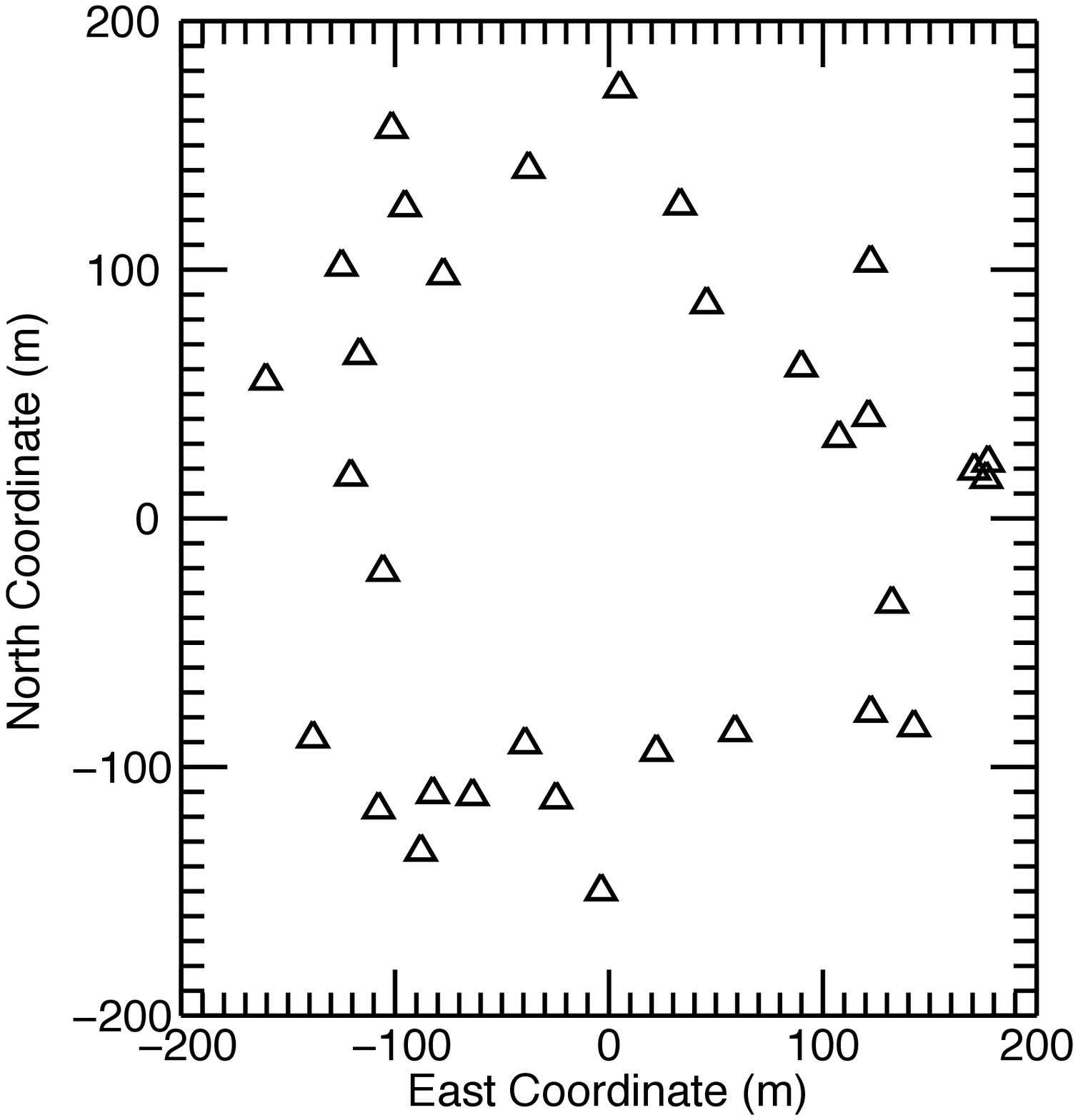}
\caption{Positions of MWA 32T tiles relative to the array centre.}
\label{antenna_pos}
\end{figure}

\begin{figure}[hbtp] 
\centering 
\includegraphics[width=.90\textwidth]{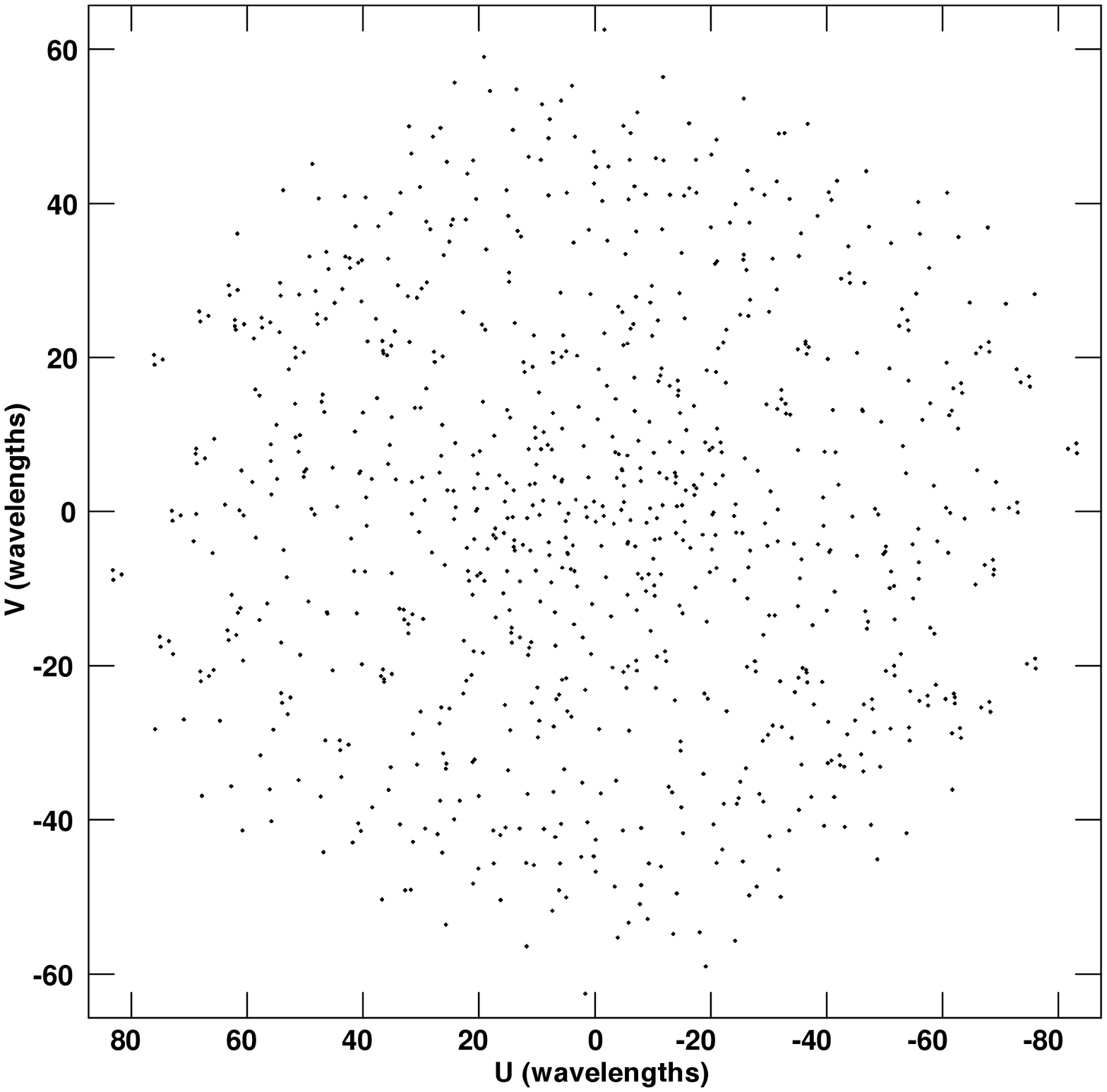}
\caption{MWA 32T instantaneous UV coverage for a pointing at the Moon at azimuth 357 degrees, elevation 52 degrees and frequency 80 MHz (wavelength 3.75 metres).}
\label{uv32T}
\end{figure}

The observations were conducted with the MWA 32T during Expedition 14 to the MWA site, which took place in September 2010. The 32 tiles of the prototype system were arranged as shown in Figure \ref{antenna_pos}. The positions are relative to the array centre. This tile configuration gives an instantaneous UV coverage as shown in Figure \ref{uv32T} for a pointing at the Moon at azimuth 357 degrees, elevation 52 degrees and frequency 80 MHz. Multi-frequency and Earth-rotation synthesis were used to maximise UV coverage for imaging. With this tile configuration the maximum baseline is approximately 350 m, giving a spatial resolution of approximately 30 arcminutes at 100 MHz. 

The Moon observations were made in a novel \textquoteleft picket fence mode\textquoteright\ which reallocated the 30.72 MHz of signal processing capacity to sample the frequency range from 80 to 200 MHz with 24 evenly spaced 1.28 MHz wide sub-bands. Four different sub-band configurations were used to fill the whole spectrum across this frequency range. The 24 sub-bands of the MWA 32T are further divided into 32 fine channels, giving a spectral resolution of 40 kHz. 

Observations of the Moon were made over the identical range of 0030-0132 LST on UTC dates 24 and 25 September 2010 (approximately 1630-1732 UTC). The resulting dataset consisted of 32, 107 second integrations (eight scans for each channel configuration) on both days. The telescope was configured such that the beam centre was aimed at the computed position of the Moon at 0100 LST on 24 September (J2000 coordinates 0\textsuperscript{h}51\textsuperscript{m}6\textsuperscript{s}, 11\degree 24$'$40$''$)  and the field allowed to drift across this fixed beam position on both days. Identical beamformer delay settings were used on both days. This was done to maximise the stability of the system as re-pointing the beam has complicated gain effects that are yet to be fully characterised.

A calibration dataset, using a setup identical to the Moon scans, but with the fixed telescope pointing centred on Fornax A, was also taken in the hour following the Moon scans on each day. These data were used for obtaining initial phase calibration solutions in order to start the self-calibration process. Fornax A could not be used to set the overall flux density scale as it has been found \citep{ord} that while phase solutions transfer well between tile pointings, the amplitudes of the complex tile gains cannot be applied to different tile pointings without being in error by several percent. Fortunately, the relatively bright source 3C33 lies in the main beam close to the Moon's position and this source was used to tie the Moon measurements to an established flux density scale.

\section{Calibration and Imaging}

The raw correlator data were averaged in time by 6 seconds and converted to UVFITS format. The dataset was then imported to the Astronomical Image Processing System (AIPS) for data reduction. The UV data were examined at the finest spectral resolution and bad channels flagged manually. Radio Frequency Interference (RFI) was sparse and generally limited to a few discrete narrow bands. Four fine channels at the edge of each sub-band, as well as the central fine channel were also flagged as these were known to be corrupted by aliasing and rounding error, respectively. Tiles 2 and 6 were also flagged during the observations as they were not behaving as expected. In order to reduce the subsequent processing time, the UV data were then averaged in frequency to give 0.64 MHz sub-bands. The complete dataset, including both days, was concatenated into one AIPS file to allow self-calibration of the dataset using the same sky model for both days. This approach tied the tile gain solutions for both days to a common reference, which was important for the Moon-sky differencing experiment.

A subset of the Fornax A data were used as an input to the AIPS \textquoteleft CALIB\textquoteright\ task to get an initial set of complex tile gain solutions for calibration. A calibration was performed on these data, assuming a sky model of a point source located at the centre. The UV range in the initial calibration was selected to exclude longer baselines so that Fornax A would not be resolved. An image was then made using the AIPS task \textquoteleft IMAGR\textquoteright, implementing Clark\textquoteright s variation of the H$\rm{\ddot{o}}$gbom CLEAN algorithm \citep{clark1980,Schwab1984}. Faceting was used to account for wide-field effects. The clean component list generated by this imaging process was then fed back into the calibration procedure as an improved sky model. This self-calibration process was repeated for several iterations using the full UV range and phase solutions only, until convergence was achieved. A final phase and amplitude self-calibration iteration was performed and the resulting image of Fornax A clearly showed a bright double source, with many fainter radio sources in the field of view.

The tile gain solutions from the final self-calibration iteration of Fornax A were then applied to the Moon data as a starting point for self-calibration. The same procedure as described above for Fornax A was used to generate an image cube using the combined set of Moon data from 24 and 25 September. Imaging was performed in 10 MHz bands to maintain a high signal to noise ratio, resulting in an image cube with 9 frequency slices, centred on the Moon's position at 0100 LST on 24 September. After the final calibration iteration was complete, the tile gain solutions were applied and the data for both days were imaged separately, producing an image cube for each day. Using the same sky model to self-calibrate the data for both days at the beginning of the process effectively tied the tile gain solutions to a common reference, allowing a fair comparison of the two days when imaged separately at the end. The motion of the Moon during the hour of observations on each day is 19 arcminutes, which is approximately the size of the synthesised beam at 150 MHz (taking the diameter of the synthesised beam as $\lambda$/350m). So at the higher end of the observed spectrum we would expect some smearing of the Moon image due to this motion.

\begin{figure*}[hbtp] 
\centering 
\subfloat
{
\includegraphics[width=.50\textwidth]{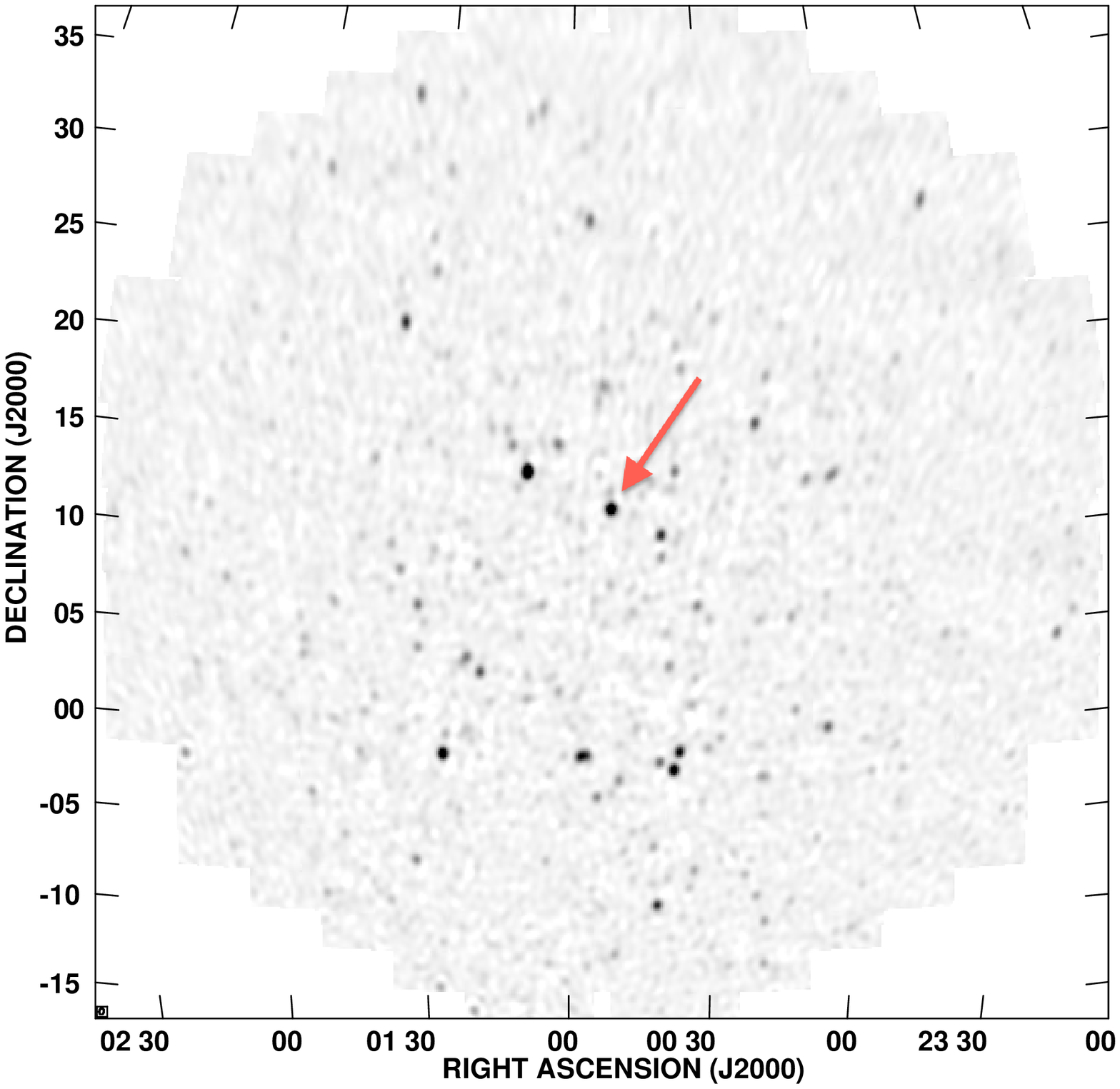}
\label{moonA24_94}
}
\subfloat
{
\includegraphics[width=.50\textwidth]{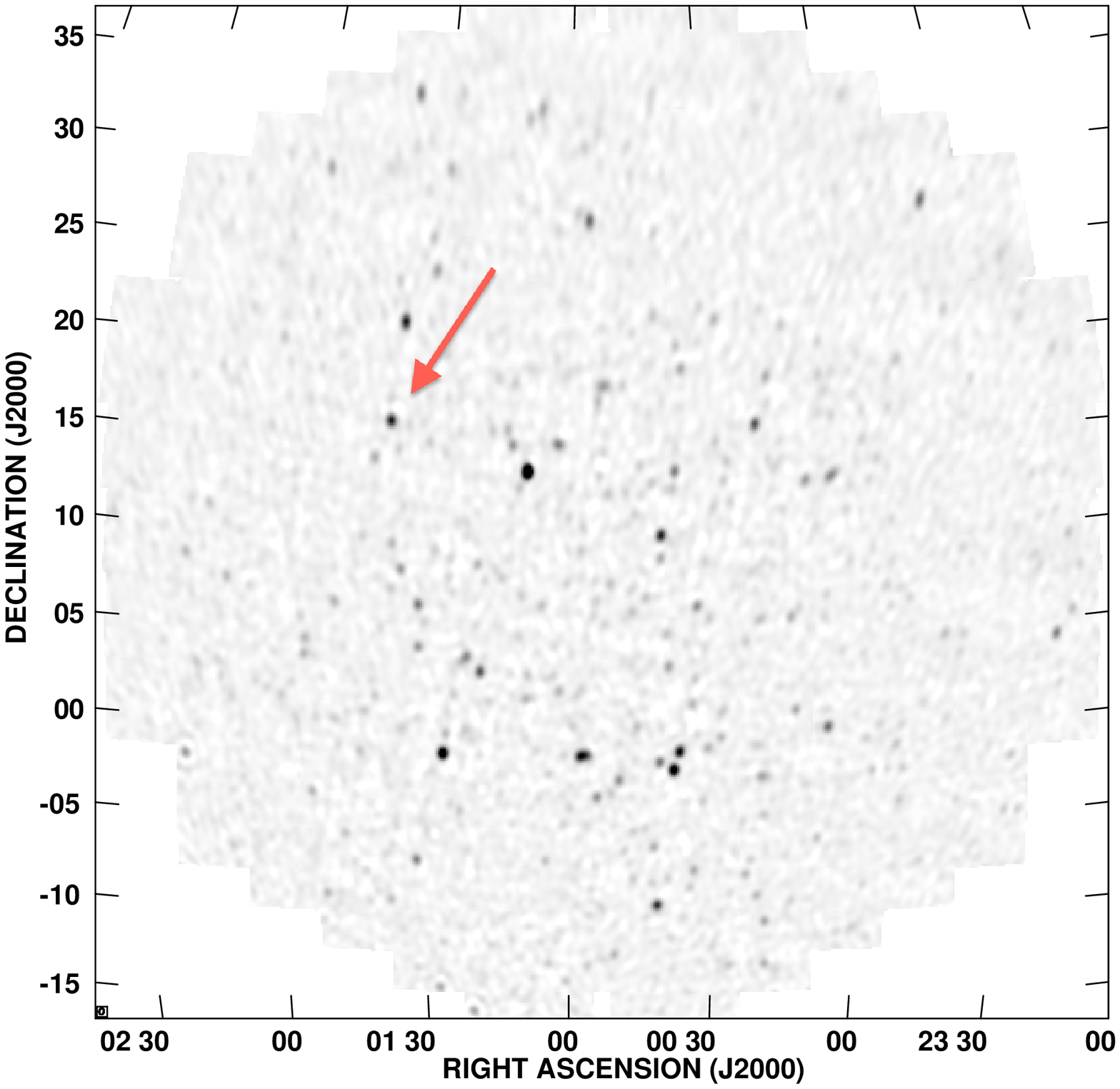}
\label{moonA25_94}
}
\caption{Moon field images on 10 MHz bandwidth, centred at 94 MHz, for 24 September 2010 (left panel) and 25 September 2010 (right panel). The flux scale is arbitrary at this stage of the calibration process and no primary beam correction has been applied. Darker greyscale indicates higher flux density.}
\label{moon24and25_94}
\end{figure*}

Examination of the cubes revealed that the Moon is present as a bright source in two of the image slices, corresponding to 94 MHz and 104 MHz, on both days. On 24 September (Figure \ref{moon24and25_94}, left panel), the bright source appears at the centre of the field and on the 25 September (Figure \ref{moon24and25_94}, right panel), the bright source has moved approximately 13 degrees across the sky as expected of the Moon. The result shown in Figure \ref{moon24and25_94} where the Moon appears to be a bright radio source rather than a cool \textquoteleft hole in the sky' does not agree with the simple prediction from Figure \ref{MoonSkyTheory}, which neglects reflected emission from the Earth.

\begin{figure}[hbtp] 
\centering 
\includegraphics[clip,trim= 100 0 100 170,angle=90]{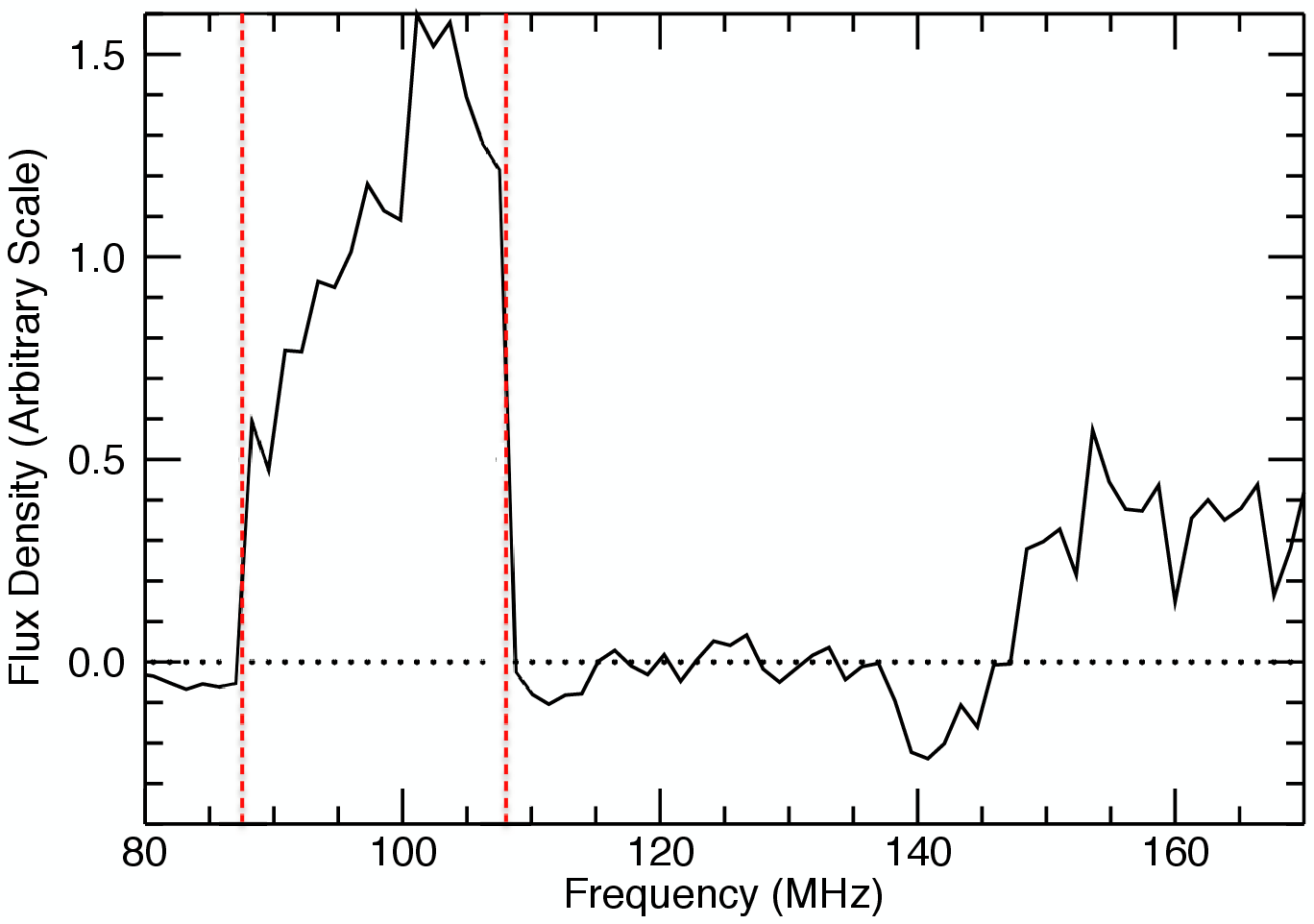}
\caption{Moon spectrum at 1.28 MHz resolution on 24 September 2010 with an arbitrary flux density scale. The vertical lines denote the FM radio band of 87.5 to 108 MHz.}
\label{ucal_moonflux}
\end{figure}

To investigate this spectral behaviour in more detail, a higher spectral resolution image cube was generated for the entire 50 degree diameter field using 1.28 MHz channels. We refer to this data cube as the Intermediate Calibration Cube, since the relative flux densities for the sources at each frequency are well determined but the overall gain passband and flux density scale have not yet been imposed. The spectrum of the integrated flux density of the central \textquoteleft Moon' pixels in the Intermediate Calibration Cube was plotted and is shown in Figure \ref{ucal_moonflux}. It is clear that the Moon's brightest disk averaged emission corresponds exactly with the internationally recognised FM radio transmission band \citep{rfspectrumplan} of 87.5 to 108 MHz. 

 \notetoeditor{These two images should appear side-by side for ease of comparison}

The imaging described up until this point has used self-calibration with an arbitrary flux density scale. Setting the flux density scale for the MWA 32T is difficult due to the currently uncharacterised behaviour of the overall gain of the system for different beamformer settings. This means that the traditional method of pointing at a known flux calibrator and then pointing at the target source cannot be used reliably and a known source within the field of view of the target is required. The relatively bright source 3C33, located approximately 4.7 degrees from the pointing centre, was therefore used as a calibration reference to establish a flux density scale for the Moon. 3C33 exhibits a power law spectrum with measured flux densities of 105.3 $\pm \ 0.4$ Jy at 74 MHz \citep{VLA_74MHz_2007} and 58.3 $\pm \ 2.9$ Jy at 178 MHz \citep{kuhr}. Interpolating these data using a power law with a single spectral index gives the following equation for the flux density of 3C33:

\begin{equation} 
S_{calc}(\nu)=S_{74} \left(\frac{\nu}{74\  \rm{MHz}}\right)^{-\alpha} \ \rm{Jy}
\label{calibeqn}
\end{equation}
where $\nu$ is the frequency in MHz,  $S_{74} = 105.3 \ \pm \ 0.4$ and $\alpha=0.67 \ \pm \ 0.06$

The spectrum of 3C33 extracted from the Intermediate Calibration Cube was measured using the AIPS task \textquoteleft ISPEC\textquoteright .\ A correction was made to this measurement to account for the drop in gain of the primary beam as a function of frequency for the source which is approximately 4.7 degrees from the beam centre. This rudimentary primary beam correction was based on a simple analytical model that didn't take into account the different beam shape of the two dipole polarisations, but is accurate for positions close the pointing centre. We refer to this corrected flux density of 3C33 as $S_{ICC}(\nu)$.

Figure \ref{ucal_moonflux} does not show the smooth spectral shape expected for a thermal spectrum. Even outside the FM band there is considerable structure above the nominal noise level. The flux density calibration specified by equation (\ref{calibeqn}) is smooth on these scales and cannot cause this sort of structure. 

\begin{figure}[hbtp] 
\centering 
\includegraphics[width=.90\textwidth]{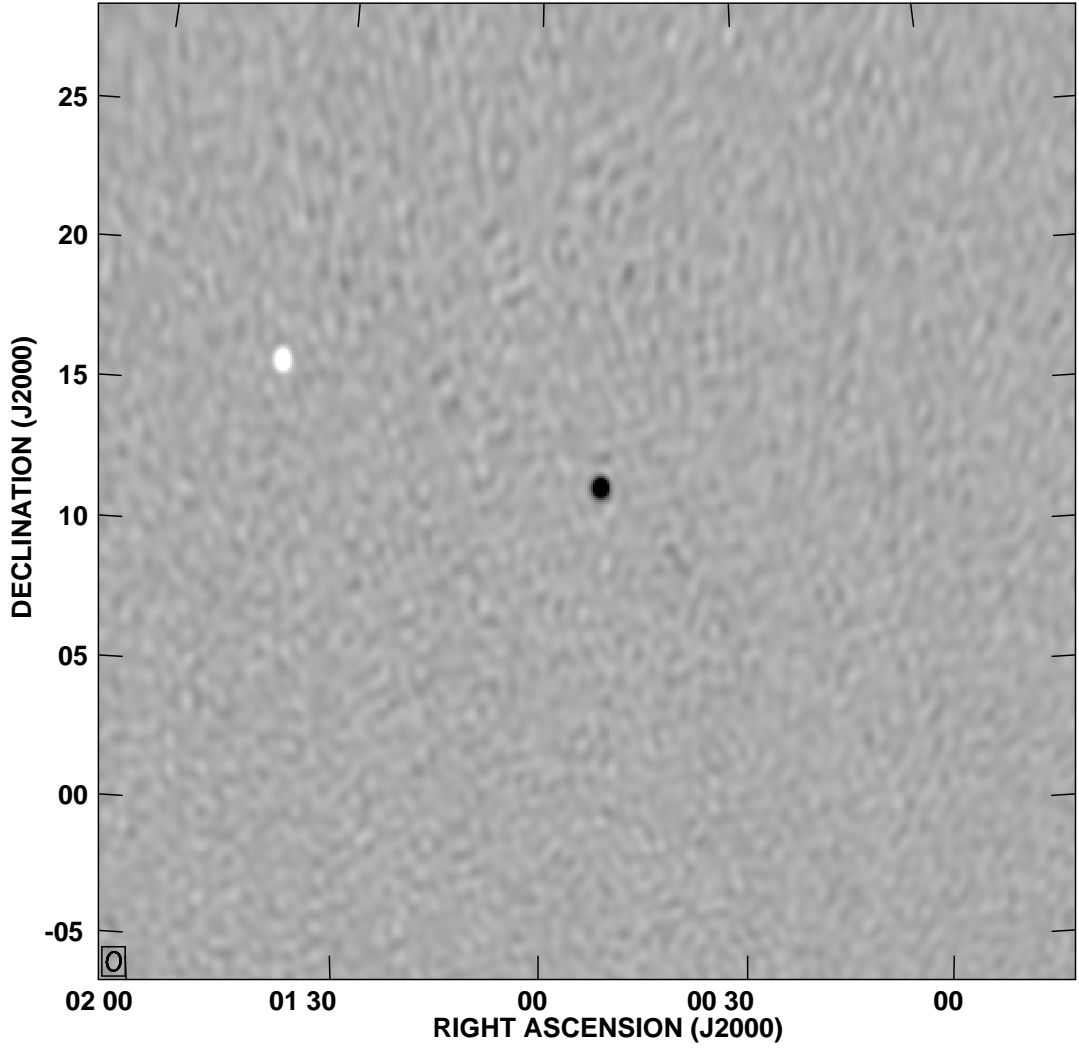}
\caption{Image slice from the Moon-sky difference cube centred at 99.835 MHz and covering 0.92 MHz of bandwidth (since 28\% of the sub-band is flagged). The flux scale of the image is arbitrary at this stage of the calibration process. Darker greyscale indicates higher (positive) flux density.}
\label{diff_im}
\end{figure}

Since the observations were made on two days with identical tracks as a function of sidereal time, certain classes of systematic imaging artefacts are expected to cancel when the {\it difference} between the two days data is examined. Specifically, the chromatic effects of residuals left from bright sources and integrated source confusion, observed with the incomplete UV coverage attained with the MWA 32T, can couple to spectral structure that will vary from beam to beam throughout the image cube. As a test of the presence of artefacts, an identical imaging and calibration process was used to produce an ICC for the 25 September, and then a difference image cube was constructed by subtracting the 25 September cube from the 24 September cube. Figure \ref{diff_im} shows an image slice from the difference cube centred at 99.835 MHz.

For a perfect instrument and perfect calibration on both days of observation with identical conditions you would expect the RMS noise in the difference image (Figure \ref{diff_im}) to approach the thermal noise level, since all sources (including low level confused sources within the synthesised beam) would be perfectly subtracted. 

The expected noise level in the Moon images, $\sigma$ in Jy, is given by:
\begin{equation} 
\sigma =10^{26}\left(\frac{2 k_{\rm{B}}T}{A_{\rm{eff}}  \epsilon_{\rm{c}}}\right) \ \sqrt{\frac{1}{N(N-1) B n_{\rm{p}} \tau}}
\label{thermal}
\end{equation}
(adapted from \citet{TMS} equation 6.62) where $ k_{\rm{B}}$ is the Boltzmann constant in J/K, $T=T_{\rm{sky}}+T_{\rm{rcv}} \approx 700 \ \rm{K}$ at 100 MHz, $A_{\rm{eff}}$ is the effective collecting area of a tile $\approx 21.5 \ \rm{m}^2$ \citep{tingay}, $N$ is the number of tiles ($N$= 30 for these observations), $\epsilon_{\rm{c}}$ is the correlator efficiency (assumed unity), $B$ is the instantaneous bandwidth in Hz, $n_{\rm{p}}$ is the number of polarisations and $\tau$ is the integration time in s. Equation (\ref{thermal}) assumes natural weighting of the visibilities and is a close approximation to the noise properties of the weighting scheme used by default in AIPS.

\begin{figure}[hbtp] 
\centering 
\includegraphics[width=1.0\textwidth]{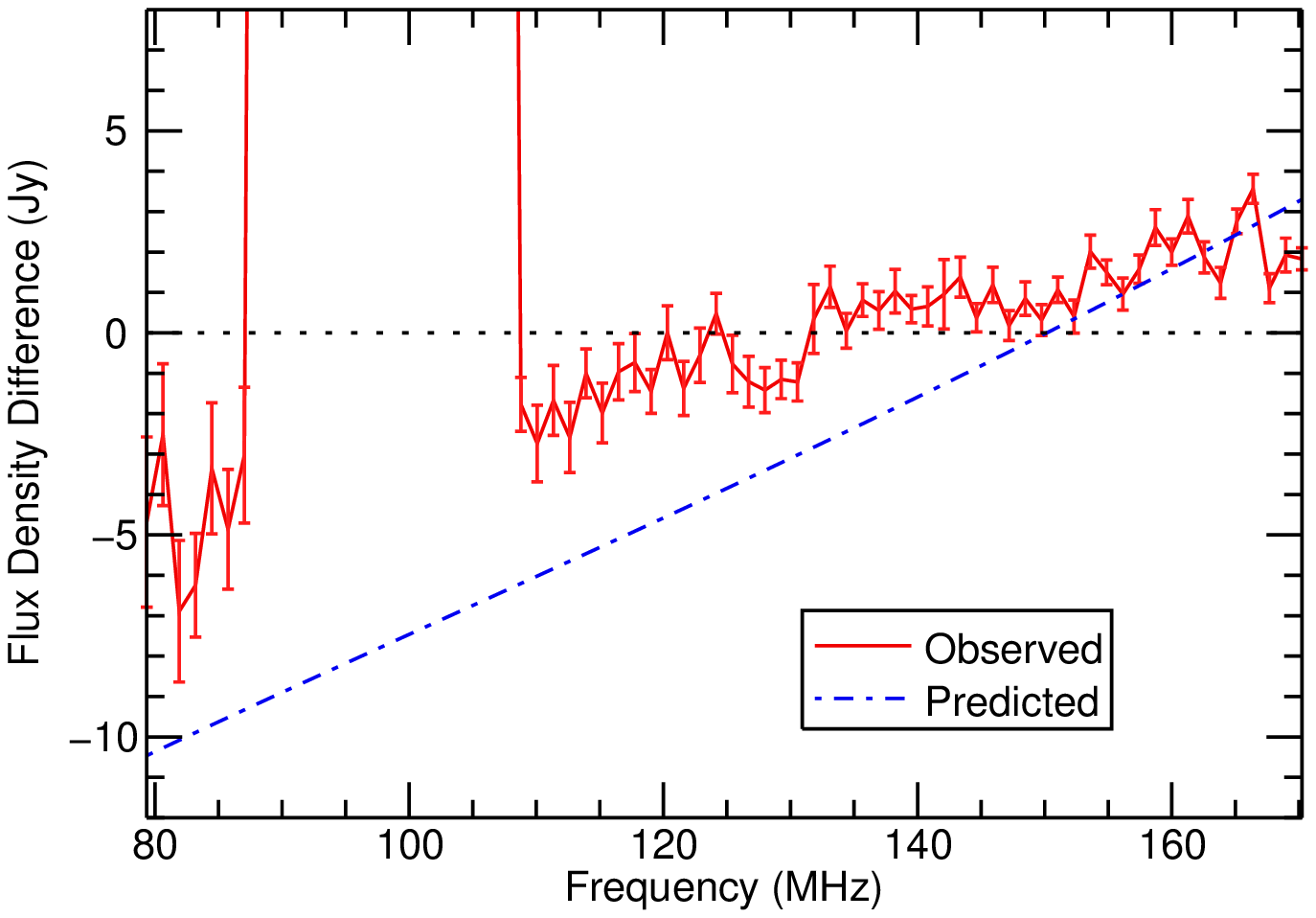}
\caption{Comparison between the observed Moon-sky flux density difference on 24 September and the difference predicted for a model of the Moon that includes thermal emission and reflection of galactic background only. The excess flux density in the observations is attributed to the reflection of the Earth's radio leakage.}
\label{moon_sky_theory_comp}
\end{figure}

We find the RMS in the difference image cube to be higher than the expected thermal noise given by equation (\ref{thermal}) by a factor of between 4 and 7 across the frequency range and that the excess is highest in the FM band. For an imaging sub-band of 0.92 MHz (since 28\% of fine channels are flagged) and an integration time of 7 minutes (the hour of observations was split into four channel configurations and half of the integration time was lost due to network bandwidth limitations) the RMS in the difference image slice shown in Figure \ref{diff_im} is approximately 1.3 Jy, whereas the expected thermal noise given by equation (\ref{thermal}) is approximately 0.2 Jy (including an additional factor of $\sqrt2$ due to the differencing). The main factor causing this discrepancy is thought to be imperfect calibration on both days causing incomplete subtraction of sources in the difference image. The same sky model was used to calibrate the concatenated data from both days. However within the FM band, there is a 60 Jy source (the Moon) in two different positions 13 degrees apart on subsequent days. This leads to a calibration solution that is slightly in error for both days and a higher than expected RMS noise in the difference cube. Another factor could be differences in the ionosphere between the two days which would affect the positions of sources in the field, resulting in imperfect subtraction. The errors in the difference image are included as error bars in Figures \ref{moon_sky_theory_comp} and \ref{moonTbzoom}.

\begin{figure}[hbtp] 
\centering 
\includegraphics[width=1.0\textwidth]{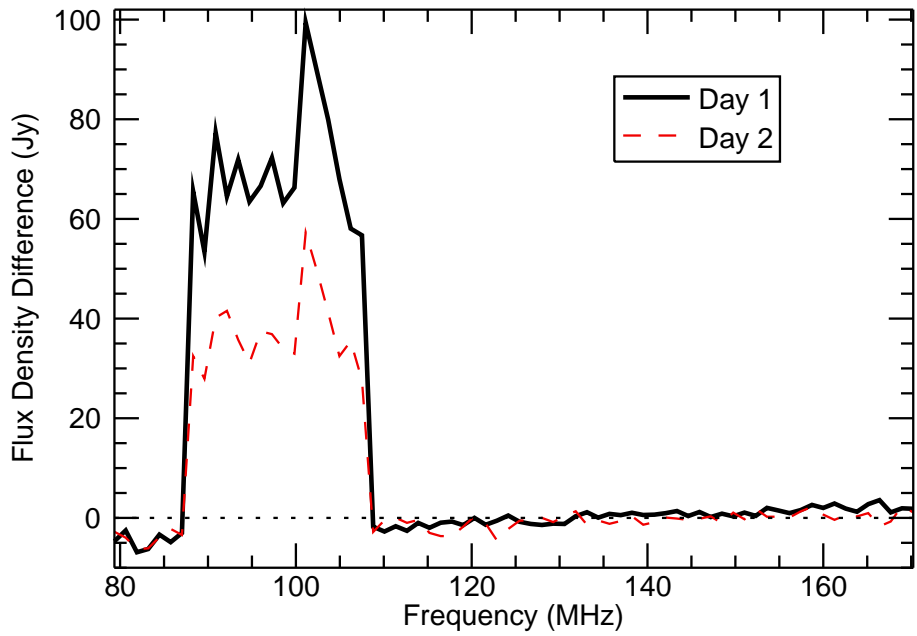}
\caption{Integrated flux density of the \textquoteleft Moon' pixels in the Moon-sky difference image on Day 1 when the Moon was at the centre of the beam and Day 2 when the Moon had drifted to approximately the half-power point of the beam.  The flux density scale has been set using 3C33 as a reference, but no primary beam correction has been applied to the Moon spectra.}
\label{diff_plot}
\end{figure}

The integrated flux density of the central pixels of the difference cube gives the difference between the Moon flux density on the 24 September and the sky flux density on 25 September. This Moon-sky difference spectrum was calibrated by application of the correction factor  $S_{calc}(\nu) / S_{ICC}(\nu)$ to give the calibrated spectrum shown in Figure \ref{diff_plot}. These observations clearly show the Moon to be a source of approximately 60 Jy across the FM radio band from 87.5 to 108 MHz and many of the residual spectral features apparent in Figure \ref{ucal_moonflux} have indeed been removed by the differencing process. The uncertainty in the flux density of 3C33 leads to an uncertainty of less than 3\% in the flux density calculated for the Moon. Following the same steps, we extracted a lower signal-to-noise ratio spectrum for the second day, when the Moon had moved approximately 13 degrees from the centre of the tile primary beam. Since the tile primary beam Full Width Half Maximum (FWHM) varies as approximately 25\textdegree /$(\nu /150 $MHz), the Moon's position on the second day corresponds approximately to the half power point of the tile primary beam at 150 MHz. This spectrum is consistent with the spectrum obtained while the Moon was at the centre of the beam as shown in Figure \ref{diff_plot}.

Excluding the FM radio band, the measured Moon-sky difference has the negative to positive trend of Figure \ref{MoonSkyTheory}, however the slope of the measured difference plot is still not a good fit to the predictions which have neglected reflected emission from Earth as shown in Figure \ref{moon_sky_theory_comp}.  The error bars in Figure \ref{moon_sky_theory_comp} represent the RMS noise in each image slice for a square region of the difference image in between the Moon's position on each of the subsequent days. It was considered whether incorrect modelling of the reflected galactic emission could account for the discrepancy between predicted and observed Moon temperatures, however the spectral index required to fit the data is approximately $\beta=4$, which is unrealistically high. Given the clear presence of reflected terrestrial FM radio transmissions from the Moon, it is likely that this discrepancy is due to additional reflected transmissions from Earth, outside of the FM radio band. 

\section{Analysis}

The strength of Earth's radio leakage signal reflected by the Moon is of interest for several reasons:
\begin{enumerate}[(i)]
\item  Reflected radio emission presents an additional challenge for experiments aiming to use the Moon as a smooth spectrum calibration source for detection of the EoR signal,
\item The Moon will appear as a transient radio source that needs to be accounted for in current and future low frequency surveys and experiments, and
\item  The strength of the detected signal allows us to calculate how bright the Earth is as a radio source in this band and hence whether it might be detectable to technologically advanced extraterrestrial civilisations.
\end{enumerate}

\begin{figure}[hbtp] 
\centering 
\includegraphics[width=1.0\textwidth]{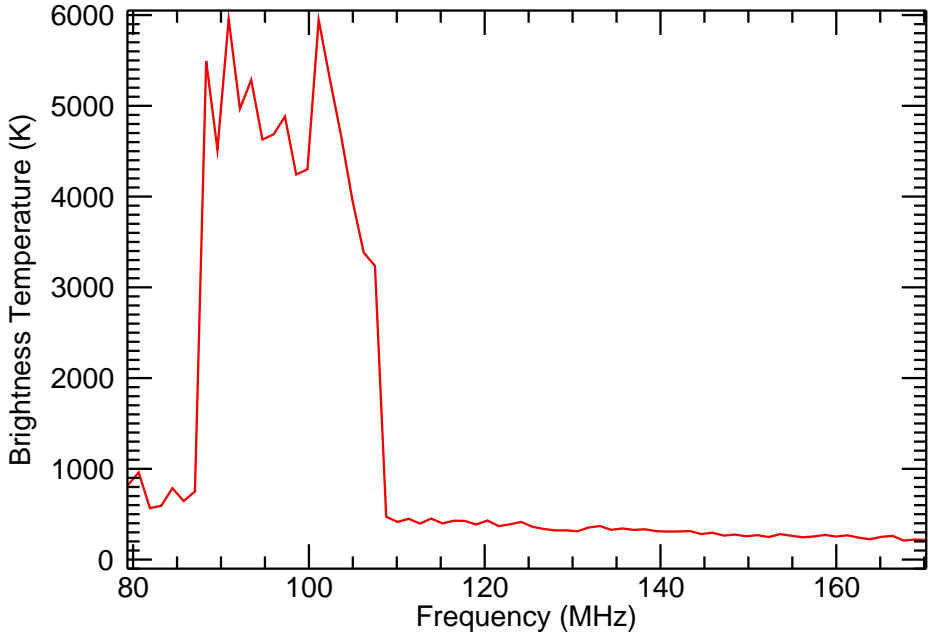}
\caption{Disk average brightness temperature of the Moon calculated from MWA 32T measurements, combined with a model of the sky temperature derived from \citet{rogersbowman2008} and \citet{,deoliveiracosta2008}. }
\label{moonTb}
\end{figure}

\begin{figure}[hbtp] 
\centering 
\includegraphics[width=1.0\textwidth]{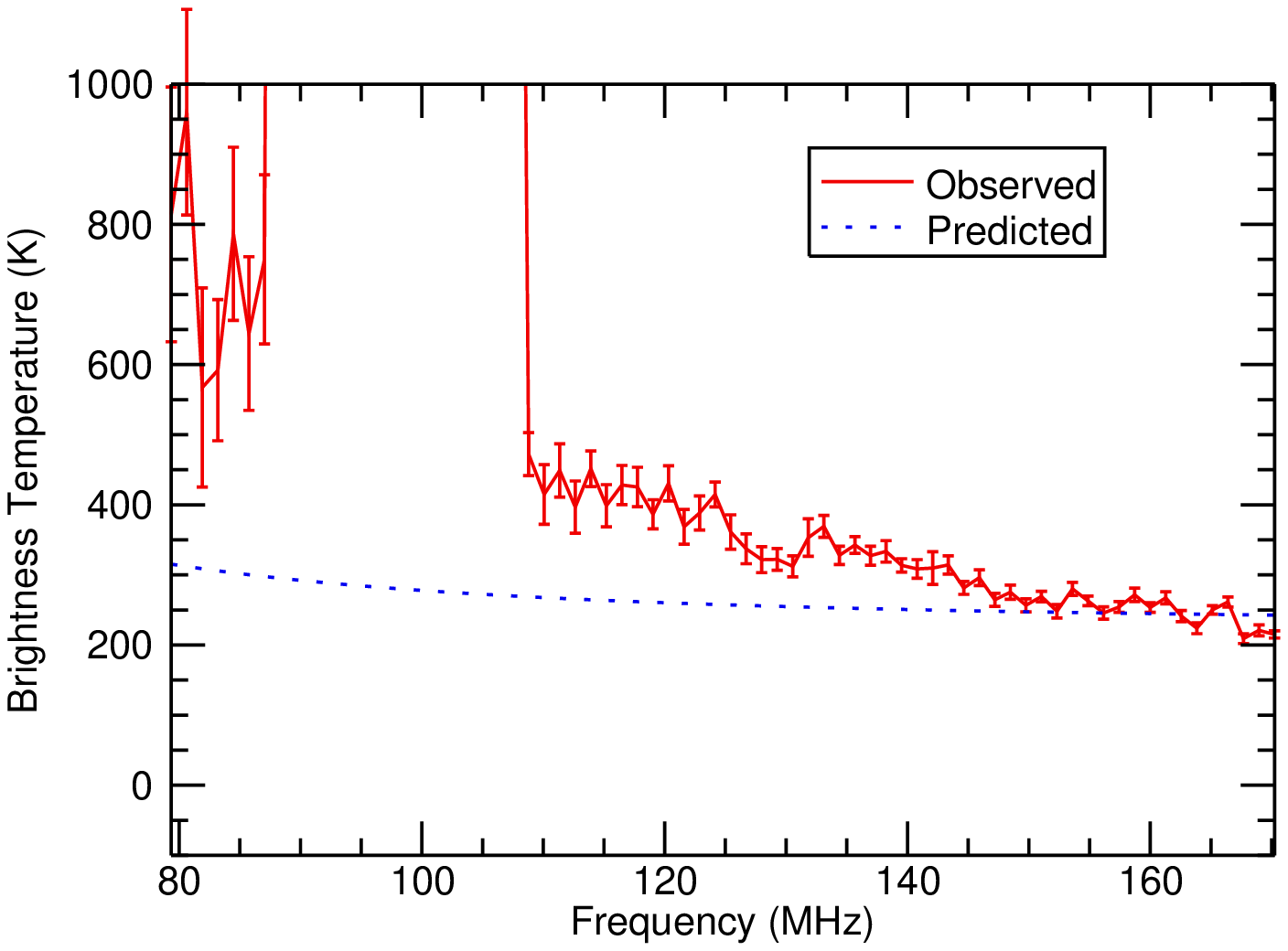}
\caption{Disk average brightness temperature of the Moon calculated from MWA 32T measurements, combined with a model of the sky temperature derived from \citet{rogersbowman2008} and \citet{,deoliveiracosta2008}. This is the same as Figure \ref{moonTb} but plotted on a different scale and with error bars based on the RMS noise of the MWA 32T difference image. The dotted line is the thermal brightness temperature of the Moon of 230K from \citet{heilesdrake63} plus the expected reflection of galactic emission. }
\label{moonTbzoom}
\end{figure}

To address point (iii) above, we can use these observations to calculate the equivalent isotropic power (EIP) of the Earth in the same way as \citet{sullivan_moon}. However, because the MWA is an interferometer, our measurements only give us the difference between the background sky flux density and the Moon's flux density. The flux corresponding to the \textquoteleft zero spacing' is missing. We overcome this by converting the observed Moon flux density to brightness temperature by rearranging equation (\ref{tempconvert}) and adding the known sky temperature given by equation (\ref{skytemp}). This gives us the observed Moon brightness temperature as shown in Figures \ref{moonTb} and \ref{moonTbzoom}. The brightness temperature of the Moon has been studied at a number of wavelengths, however the most relevant to these observations is that of \citet{heilesdrake63} which shows that we should expect the Moon to have a constant brightness temperature of 230 K at these frequencies. The excess brightness temperature due to reflected emission from the Earth is therefore calculated by subtracting the Moon's thermal emission and the component due to reflected galactic emission from this spectrum. This excess brightness temperature can then be converted back to flux density $S_{m}$ by equation (\ref{tempconvert}) and to an EIP in Watts by:
\begin{equation} 
P_{earth}=\frac{4^{2} \pi  10^{-26} d_{m}^{4} \Delta\nu S_{m}}{r_{m}^{2} \rho_{m}}
\label{Pearth_eqn}
\end{equation}
where $d_{m}$ is the average Earth-Moon distance in m, $\Delta\nu$ is the size of the frequency bins in Hz, $S_{m}$ is the observed flux density of the Moon in Jy,  $r_{m}$ is the radius of the Moon in m, and $ \rho_{m}$ is the Moon's reflectivity of 0.07 \citep{evans_moon}.

\begin{figure}[hbtp] 
\centering 
\includegraphics[width=1.0\textwidth]{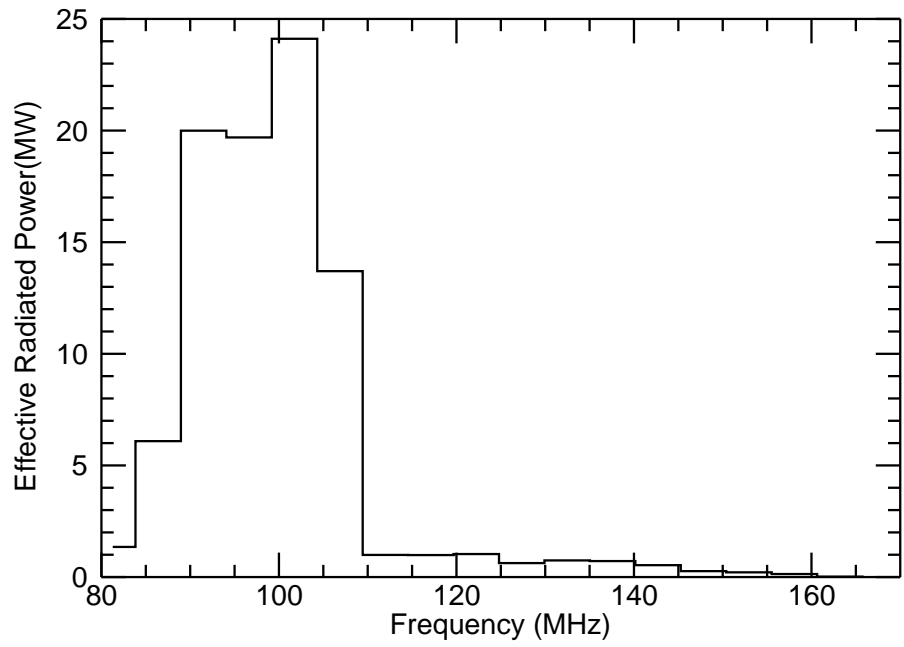}
\caption{The equivalent isotropic power (EIP) of radio emissions from Earth as measured from MWA 32T observations of the Moon. The plot is a histogram with 5 MHz bins.}
\label{erp}
\end{figure}

Evaluating equation (\ref{Pearth_eqn}) for the Moon's spectrum using 5 MHz wide frequency bins gives the histogram in Figure \ref{erp}. 
To understand what the signal shown in Figure \ref{erp} means from the perspective of the Search for Extraterrestrial Intelligence (SETI),  we can calculate how bright the Earth might appear to an extraterrestrial observer. Summing the EIP over the 20.5 MHz of the FM band gives a value of approximately 77 MW, so to an observer orbiting our closest neighbouring star, Proxima Centauri, the Earth would appear to be an extremely faint source of approximately 19 nJy within the FM radio band. As a comparison, the quiet Sun is a source of approximately 60,000 Jy \citep{lantos} as viewed from the Earth in this frequency band during periods of low solar activity. The Sun's total EIP in the FM band is therefore approximately 3.5 GW, so viewed from Proxima Centauri it would appear as a 0.85 $\mu$Jy source and greatly outshine the Earth. 

Detection of such a small signal by a nearby, technologically advanced civilisation would require far superior instrumentation to our own, with even greater collecting area than is expected for future telescopes such as the Square Kilometre Array (SKA). For a three sigma detection at these frequencies with an SKA equivalent instrument with baselines long enough to resolve the Earth from the Sun, the integration time required is on the order of hundreds of years. This is due largely to the high background sky temperature which dominates the system temperature at low frequencies. \citet{loeb2007} found that beamed radio leakage from a civilisation within 23 pc of Earth, using high power military type radars similar to our own, should be detectable by the next generation of low frequency radio interferometers. However we note that most of these types of systems on Earth operate at higher frequencies. For this reason, radio leakage from the Earth at higher frequencies is more likely to be detectable, as the background sky temperature is lower and our highly beamed, military radars operate in the GHz frequency range. SETI experiments planned for the SKA are therefore more focussed at frequencies between 1 and 10 GHz \citep{tarter_SETI}. The long baselines of the SKA will also provide the required angular resolution and RFI rejection capabilities required for such SETI searches, the importance of which have been demonstrated by recent proof-of-concept SETI observations using Very Long Baseline Interferometry (VLBI) techniques \citep{ramp}.

The observed Moon-sky flux density difference indicates that more observations are needed to properly characterise the Moon's radio properties at low frequencies if the Moon is to be used as a smooth spectrum calibrator for detection of the EoR signal. The Very High Frequency (VHF) part of the radiofrequency spectrum from 30 to 300 MHz is used for many applications \citep{rfspectrumplan} including FM broadcasting, aeronautical radionavigation and communications and amateur radio to name a few. These will all contribute to the reflected radio signal received on Earth from the Moon as evidenced by Figure \ref{moon_sky_theory_comp}. However there are methods that could be employed to reduce the impact of this reflected RFI on observations and some of these are discussed below.

The RFI signals from the Moon are expected to be strongest from the sub-Earth point where the surface behaviour is closest to specular reflection. The higher spatial resolution of the completed MWA with 128 tiles and 3 km baselines will enable us to isolate the specular and diffuse reflection regions and take spectra from regions of the Moon less affected by reflected RFI from Earth.

A higher spectral resolution investigation may also provide a means of avoiding reflected RFI. In many countries including Australia and the United States \citep{itu} FM broadcast frequencies are assigned with a channel separation of 200 kHz and a maximum frequency deviation of $\pm 75$ kHz, leaving a \textquoteleft guard band' of at least 25 kHz above and below the modulation.  Observation in these guard bands should be free from RFI contamination or at least be affected at a reduced level. Similarly radio-quiet bands may be found in regions of the spectrum outside of the FM band if observing at higher spectral resolution.

The reflected radio emission from the Moon will clearly also vary strongly as a function of time as different regions of the Earth come into the line-of-sight of the Moon. Since terrestrial broadcasting is generally directed toward the horizon rather than straight up, the worst RFI contamination is likely to occur when the Moon's position falls near the plane of the horizon at the most highly populated regions of the Earth. Hence RFI contamination may be reduced by planning observations at times when the geometry between the Moon, the telescope and population centres on Earth minimises the radio signal beamed toward the Moon. 
 
An additional concern is that the occasional presence of the Moon in observations of deep fields that are visited repeatedly may present complications due to the rapid movement of the Moon with its strong reflected signal, which itself is rich in time-variable spectral structure as the Earth rotates and different population centres illuminate the Moon with radio transmissions.  Furthermore, the Moon subtends sufficient solid angle that ensembles of background sources will be steadily occulted and re-exposed in areas of long integration. This is a concern for some instruments, since the principles of aperture synthesis and coherent integration with interferometers rely on the constancy of the radio sky throughout the integration. However, instruments such as the MWA and LEDA \citep{greenhill2012} will implement snapshot imaging with subsequent combination in the image domain, an approach that might be made robust to this kind of variability.

\section{Conclusions}

The MWA 32 tile prototype system was used to observe the Moon over two subsequent nights in order to test the viability of using the Moon as a smooth spectrum calibration source for EoR signature detection. A direct differencing experiment between the Moon and sky flux density from the same patch of sky over the two nights found that the Moon does exhibit the properties expected for a cool thermally emitting body, but that the spectrum is corrupted by radio leakage from Earth, most notably within the FM radio band of 87.5 to 108 MHz. From this reflected signal we were able to measure the EIP of the Earth in the FM radio band which was found to be approximately 77 MW.

The implications of these observations are that RFI from Earth will be a significant challenge to overcome if the Moon is to be used as a smooth spectrum calibration source for EoR detection. Reflected terrestrial radio emissions will have to be carefully taken into account in any future experiments and more Moon observations are required to characterise this RFI.  Additionally, it was found that the total signal being transmitted from the Earth in the FM radio band would be difficult to detect, even by a nearby technologically advanced civilisation with superior radio astronomy technology to our own.

\section{Acknowledgments}

This work uses data obtained from the Murchison Radio-astronomy Observatory. We acknowledge the Wajarri Yamatji people as the traditional owners of the Observatory site. Support came from the U.S. National Science Foundation (grants AST-0457585, AST-0908884 and PHY-0835713), the Australian Research Council (grants LE0775621 and LE0882938), the Centre for All-sky Astrophysics (CAASTRO - an Australian Research Council Centre of Excellence, funded by grant CE110001020), the U.S. Air Force Office of Scientific Research (grant FA9550-0510247), the Smithsonian Astrophysical Observatory, the MIT School of Science, the Raman Research Institute, the Australian National University, the iVEC Petabyte Data Store, the Initiative in Innovative Computing and NVIDIA sponsored Center for Excellence at Harvard, and the International Centre for Radio Astronomy Research, a Joint Venture of Curtin University of Technology and The University of Western Australia, funded by the Western Australian State government. Funding support for the MWA project has also been provided by the Australian Federal government via the National Collaborative Research Infrastructure Strategy and Astronomy Australia Limited, under contract to Curtin University of Technology.

{\it Facilities:} \facility{MWA}.



\begin{thebibliography}{}

\bibitem[ACMA(2009)]{rfspectrumplan} Australian Communications and Media Authority, 2009, Australian Radiofrequency Spectrum Plan 2009 (Melbourne, VIC: ACMA), http://www.acma.gov.au

\bibitem[Bernardi et al(2009)]{berardi2009} Bernardi, G., de Bruyn, A. G., Brentjens, M. A., et al, 2009, A\&A, 500, 965

\bibitem[Bowman \& Rogers(2010)]{bowmanrogers2009}Bowman, J. D., Rogers, A. E. E., 2010, Nature, 468, 796

\bibitem[Clark(1980)]{clark1980} Clark, B., 1980, A\&A. 89, 377

\bibitem[de Oliveira-Costa et al(2008)]{deoliveiracosta2008} de Oliveira-Costa, A., Tegmark, M., Gaensler, B. M., 2008, MNRAS, 247

\bibitem[DeWitt \& Stodola(1949)]{projdiana} DeWitt, J. H. Jr., \& Stodola, E., K., 1949, Proc Inst Rad Eng, 37, 229

\bibitem[Evans(1969)]{evans_moon} Evans, J. V., 1969, ARA\&A, 7, 201

\bibitem[Furlanetto et al(2006)]{furlanetto2006} Furlanetto, S. R., Oh, S. P., Briggs, F. H., 2006, Physics Reports, 433, 4-6

\bibitem[Greenhill \& Bernardi(2012)]{greenhill2012} Greenhill, L. J., Bernardi, G., 2012, arXiv1201.1700G

\bibitem[Hagfors et al(1969)]{hagfors69} Hagfors, T., Green, J. L., Guill\'{e}n, A., 1969, AJ, 74, 10

\bibitem[Heiles \& Drake(1963)]{heilesdrake63} Heiles, C. \& Drake F. D., 1963, ICARUS, 2, 281-292

\bibitem[ITU(2001)]{itu} International Telecommunication Union, 2001, Transmission standards for FM sound broadcasting at VHF, Recommendation BS.450-3 (Geneva, Switzerland: ITU), http://www.itu.int

\bibitem[Kassim et al(2007)]{VLA_74MHz_2007} Kassim, N. E., Joseph, T., Lazio, W., et al, 2007, ApJS, 172, 686

\bibitem[Katz \& Franco(2011)]{katz} Katz, A., \& Franco, M., 2011, IEEE Microwave Magazine, 12, 62

\bibitem[K\"{u}hr et al(1981)]{kuhr}K\"{u}hr, H., Witzel, A., Pauliny-Toth, I. I. K., Nauber, U.,  1981, A\&AS, 45, 367

\bibitem[Landecker \& Wielebinski(1970)]{landecker1970} Landecker, T. L. \& Wielebinski, R., 1970, AuJPA, 16, 1

\bibitem[Lantos \& Avignon(1975)]{lantos} Lantos, P. \& Avignon, Y., 1975, A\&A, 41, 137

\bibitem[Loeb \& Zaldarriaga(2007)]{loeb2007} Loeb, A. \& Zaldarriaga, M., 2007, JCAP, 1, 20

\bibitem[Lonsdale et al(2009)]{mwadesign} Lonsdale, C. J., Cappallo, R. J., Morales, M. F., et al, 2009, in Proc. IEEE, 97, 8

\bibitem[Morales \& Wyithe(2010)]{moraleswyithe2010} Morales, M. F., Wyithe, J. S. B., 2010, ARA\&A, 48, 127

\bibitem[Oberoi et al(2011)]{oberoi} Oberoi, D., Matthews, L. D., Cairns, I. H., 2011, ApJ, 728, 270

\bibitem[Ord et al(2010)]{ord} Ord, S. M., Mitchell, D. A., Wayth, R. B., et al, 2010, PASP, 122, 1353

\bibitem[Pritchard \& Loeb(2011)]{pritchard_loeb} Pritchard, J. R., \& Loeb, A. 2011, arXiv.1109.6012

\bibitem[Rampadarath et al(2012)]{ramp} Rampadarath, H., Morgan, J. S., Tingay, S. J., Trott, C. M., 2012, AJ, 144, 38

\bibitem[Rogers \& Bowman(2008)]{rogersbowman2008} Rogers, A. E. E., Bowman, J., D.,  2008, AJ, 136, 641

\bibitem[Schwab(1984)]{Schwab1984} Schwab F. R., 1984, AJ, 89, 7

\bibitem[Shaver et al(1999)]{shaver_etal_1999} Shaver, P. A., Windhorst, R. A., Madau, P., de Bruyn, A. G., 1999, A\&A, 345, 380

\bibitem[Sullivan et al(1978)]{sullivan1978} Sullivan, W. T., Brown, S., Wetherill, C., 1978, Sci, 199, 377

\bibitem[Sullivan \& Knowles(1985)]{sullivan_moon} Sullivan, W. T., \& Knowles, S. H., 1985, IAUS, 112, 327

\bibitem[Tarter(2004)]{tarter_SETI} Tarter, J. C., 2004, NewAR, 48, 1543

\bibitem[Thompson et al(2001)]{TMS}Thompson, A. R, Moran, J. M., Swenson, G, W. Jr., 2001, Interferometry and Synthesis in Radio Astronomy (2nd ed.; New York:Wiley)

\bibitem[Tingay et al(2012, in press)]{tingay} Tingay, S. et al, 2012, PASA, in press (arXiv:astro-ph/1206.6945)

\bibitem[Williams et al(2012)]{williams} Williams, C. L., Hewitt, J. N., Levine, A. M., et al,  2012, ApJ, 755, 47

\end{thebibliography}
\end{document}